\documentstyle[aps,prl,twocolumn]{revtex}

\begin{document}


\par
\begingroup
\twocolumn[%
\vspace*{-3ex}
\hspace*{\fill}{LANCASTER-TH/9502}\\
\hspace*{\fill}{hep-ph/9502417}\hspace*{3.5em}
\vskip 10pt
{\large\bf\centering\ignorespaces
Cosmology with a TeV mass GUT Higgs
\vskip2.5pt}
{\dimen0=-\prevdepth \advance\dimen0 by23pt
\nointerlineskip \rm\centering
\vrule height\dimen0 width0pt\relax\ignorespaces
David H.~Lyth and Ewan D.~Stewart
\par}
{\small\it\centering\ignorespaces
School of Physics and Chemistry, Lancaster University,
Lancaster LA1 4YB,~~~U.~K.
\par}
{\small\rm\centering(\ignorespaces\today\unskip)\par}
\par
\bgroup
\leftskip=0.10753\textwidth \rightskip\leftskip
\dimen0=-\prevdepth \advance\dimen0 by17.5pt \nointerlineskip
\small\vrule width 0pt height\dimen0 \relax
The most natural way to break the GUT gauge symmetry is with a Higgs field
whose vacuum expectation value is of order $10^{16}\,\mbox{GeV}$ but
whose mass is of order $10^2$ to $10^3\,\mbox{GeV}$. This can lead to a
cosmological history radically different from what is usually assumed to
have occurred between the standard inflationary and nucleosynthesis epochs,
which may solve the gravitino and Polonyi/moduli problems in a natural way.
\par\egroup
\vskip2pc]
\thispagestyle{plain}
\endgroup


It is generally thought that the fundamental interactions respect local
supersymmetry, called supergravity \cite{supergravity}.
The most popular implementations of supergravity generally encounter two
cosmological problems.
One is an over-abundance of the gravitino, the spin $3/2$ superpartner of
the graviton \cite{gravitino,Fischler}.
The other is an over-abundance of one or more species of spin zero
particle, with mass $m_\Phi \sim 10^2$ to $10^3\,\mbox{GeV}$ and
gravitational strength interactions
\cite{Polonyi,dilaton,Banks,Randall,bertbento,Steinhardt}.
The latter problem was first recognised \cite{Polonyi}
in an early model of supergravity involving the Polonyi field,
and became known as the Polonyi problem.
It has persisted in versions of supergravity derived from the superstring
\cite{dilaton}, where the troublesome particles are the moduli
generic to such theories. We will use the term `moduli' to cover
all cases, and for simplicity consider only one species corresponding to
a real field $\Phi$.

The observed ratios of the three gauge couplings of the Standard Model
suggest that the correct supergravity model will contain a GUT
(Grand Unified Theory), with a unification scale
$M_{\rm GUT} \sim 10^{16}\,\mbox{GeV}$.
The GUT is broken down to the Standard Model when a scalar field $h$,
charged under the GUT gauge symmetry, but neutral under the Standard Model
gauge symmetry, and called the GUT Higgs, acquires an expectation value
$ |h| \equiv M_{\rm GUT} $.
This was originally supposed to be achieved by a scaled-up version of the
Standard Model Higgs potential,
$ V = \lambda ( |h|^2 - M_{\rm GUT}^2 )^2 $ with $ \lambda \sim 1 $.
In that case the energy scale $V_0^{1/4}$ set by the height
$ V_0 \equiv V(0) $ is of order $M_{\rm GUT}$, and the mass of the
GUT Higgs is also of order $M_{\rm GUT}$.
Such a potential leads to a history of the universe that has been widely
described \cite{inflation,vs}.
But in the context of supergravity and superstrings, where one hopes
to generate all energy scales dynamically in terms of the Planck mass
$m_{\rm Pl}$, a potential of this kind does not seem very likely.
It is more natural \cite{GUT} to suppose that $|h|$
corresponds to a direction in field space which is
exceptionally or absolutely flat, before the non-perturbative
effects that lead to supersymmetry breaking are taken into account.
After supersymmetry breaking the potential is of the form
\begin{equation}
\label{V}
V = V_0 - \frac{1}{2} m^2 |h|^2 +...
\end{equation}
with $m \sim 10^2$ to $10^3\,\mbox{GeV}$
(the scale of supersymmetry breaking). The higher order
terms, which still correspond to an exceptionally flat direction,
are negligible for $|h| \ll M_{\rm GUT}$, but generate a minimum
at the required value $|h|=M_{\rm GUT}$. The mass of the GUT Higgs is now
only of order $m$, and the height $V_0^{1/4}$ is only of order
$(m M_{\rm GUT})^{1/2} \sim 10^9$ to $10^{9.5}\,\mbox{GeV}$.

The purpose of this paper is to point out that such a flat GUT potential
may imply a history of the early universe very different from
the usual one, in which the gravitino and moduli problems may be solved.
Some aspects of this history have been considered by previous authors
\cite{decay,yam,therm,Ross,yam2,interm}, but they did not consider the
effect of what we shall call Thermal Inflation. Indeed, as far as we can
discover the entire previous literature on this type of inflation
consists of precisely one sentence \cite{therm}.

The history is summarized in the Table. It begins as usual with an era of
ordinary inflation \cite{inflation} in which the energy density
$\rho$ is dominated by the potential $V$ of the scalar fields, with one of
them, termed the inflaton, slowly rolling down it.
The potential at the end of ordinary inflation, $V_{\rm inf}$, must satisfy
$ V_{\rm inf}^{1/4} \lesssim 10^{16}\,{\rm GeV} $
to avoid generating too much large scale cmb anisotropy \cite{bound}.
Of the many models of this era that have been proposed,
the only ones that are sensible in the context of supergravity
are Natural Inflation \cite{natural}, and some versions
\cite{fvi,Ewan} of Hybrid Inflation \cite{hybrid}.
In none of them is the inflaton a Higgs field.

During ordinary inflation, non-inflaton fields typically acquire masses
squared at least of order $H^2$ \cite{displacement,fvi},
which may be of either sign \cite{Ross}.
We make the assumption that the effective GUT Higgs mass squared is
positive during inflation, so that it is trapped at $|h|=0$.
At some epoch after ordinary inflation `reheating' occurs,
which means that the bulk of the energy density thermalizes at some
`reheat temperature' $T_{\rm R} \sim (\rho/g_\ast)^{1/4}$,
where $g_\ast$ is the effective number of massless species.
For simplicity, we assume in what follows that
$ T_{\rm R} \gtrsim V_0^{1/4} $.
When the GUT Higgs field is in thermal equilibrium at a temperature in
excess of some critical value $T_{\rm c}\sim m$, its effective
potential acquires a minimum at $|h|=0$ \cite{yam,tomislav}.
Even if full reheating is long delayed, one expects some fraction
$\epsilon$ of the energy density to thermalize promptly leading to an
initial temperature
$ T_{\rm inf} \sim ( \epsilon V_{\rm inf} / g_\ast )^{1/4} $.
Provided that
$ \epsilon$ exceeds $ ( T_{\rm c} / T_{\rm R} )^{4}
	( V_{\rm inf}^{1/4} / T_{\rm R} )^{4/3}\lesssim 10^{-16} $
one will have $ T > T_{\rm c} $ even before full reheating,
and we assume that this is so.
The net effect of these conditions is to
trap the GUT Higgs at $|h|=0$ until $T=T_{\rm c}$.

Thermal Inflation begins at $ T \sim ( V_0 / g_\ast )^{1/4} $,
when the GUT potential $V_0$ starts to dominate the thermal
energy density $ \sim g_\ast T^4 $, and it ends at $T=T_{\rm c}$,
after $ \ln[(V_0/g_\ast)^{1/4}/T_{\rm c}] \sim 15 $ $e$-folds
of inflation, when $|h|$ rolls away from zero.
At around this same temperature the Standard Model Higgs also rolls
away from zero. Thus the full GUT symmetry breaks more or less
directly to the broken Standard Model symmetry
$ SU(3)_{\rm C} \otimes U(1)_{\rm EM}$ at $T \sim m$.
(Note that the expansion of the universe
does not prevent this phase transition because the Hubble time
$H^{-1}\sim (m_{\rm Pl}/M_{\rm GUT})m^{-1}$ is bigger than the
duration $\sim m^{-1}$ of the transition.
We define the Planck mass as
$m_{\rm Pl} = (8\pi G)^{-1/2} = 2.4 \times 10^{18}\,\mbox{GeV}$.)
This is in contrast with the traditional case of a non-flat potential,
where the GUT symmetry breaks to the unbroken Standard model symmetry at
$T \sim M_{\rm GUT}$, leaving the electroweak phase transition to
complete the breaking at $T \sim m$.

After Thermal Inflation ends, relic radiation from the first
Hot Big Bang plays no further role and in particular the
quark-hadron transition is of no interest.
A Cold Big Bang now begins, with $\rho$ dominated by the oscillation
of the homogeneous GUT Higgs field,
or equivalently by non-relativistic GUT Higgs particles (matter).
After a few Hubble times the amplitude of the oscillation has been
reduced by the expansion of the universe, so that the GUT Higgs field
is of order $M_{\rm GUT}$.
This means that the GUT Higgs couples directly only to particles with
mass of order $M_{\rm GUT}$, so that its interaction with ordinary
particles is very weak.

The Cold Big Bang ends at a time of order the inverse GUT Higgs decay
rate,
$M_{\rm GUT}^2/m^3$ \cite{decay} (note that parametric resonance
effects \cite{param} are unlikely to be important, since the decay rate is
much less than the mass).
If the decay products thermalize the temperature is then
$ T_{\rm decay} \sim m_{\rm Pl}^{1/2} m^{3/2} / M_{\rm GUT}
\sim 10\,{\rm MeV}$ to $100\,{\rm keV}$.
In order not to affect nucleosynthesis one will need $T_{\rm decay}$
at the upper end of this range, which among other things ensures
thermalization (except for the LSP which we discuss later).
This formally corresponds to $m \simeq 10^3\,\mbox{GeV}$ but the
uncertainties in our estimates are such that a value
$m \simeq 10^2\,\mbox{GeV}$ cannot be excluded.

Now let us ask about dangerous relics.
In the usual cosmology, entropy conservation is a good approximation
and as a result the entropy density $s \sim g_\ast T^3$ and the number
density $n$ of any stable relic have a constant ratio after the relic
stops interacting (`freezes out'). The GUT Higgs decay releases a huge
amount of entropy, increasing it by a factor
$ \Delta \sim 3 g_\ast^{-1} V_0 T_{\rm decay}^{-1} T_{\rm c}^{-3} $.
(In this expression $g_\ast$ refers to the unbroken
GUT at $T \sim T_{\rm c}$, and from now on we replace it by the estimate
$g_\ast/3\sim 10^2$.)
The present number density of any species created before that time
is diluted by this factor, if its initial number density depends only
on the temperature.
Setting $ m = T_{\rm c} = 10^2 $ to $10^3\,\mbox{GeV}$ gives
$ \Delta \sim 10^{29}$ to $10^{30}$.
Note that because of Thermal Inflation and the Cold Big Bang,
a given scale leaving the horizon during ordinary inflation does so
$ (1/3) \ln \Delta \sim 23 $ $e$-folds later than in the usual cosmology,
which could significantly affect predictions of the spectral indices of
the perturbations produced during ordinary inflation.

The gravitino is harmless if $ n_{3/2}/s \lesssim 10^{-12} $ to $10^{-15}$
at nucleosynthesis \cite{constraint}. Gravitinos created during the first
Hot Big Bang have an abundance no bigger than the thermal equilibrium
value $ n_{3/2}/s \sim 10^{-3}$ so their present abundance is
far inside the above bound. Gravitinos are not produced by the GUT Higgs
decay if $ m_h < m_{3/2} $. Finally,
gravitinos generated during the second Hot Big Bang are harmless,
because the relevant bound $T \lesssim 10^5\,\mbox{GeV}$ is amply satisfied
\cite{Fischler}. (Note that this is five orders of magnitude stronger
than earlier estimates, which neglected an important mechanism for
creating gravitinos.)

Moduli are also harmless if $ n_\Phi / s \lesssim 10^{-12} $ to $10^{-15}$
at nucleosynthesis \cite{constraint}.
We will take $\Phi=0$ to be the vacuum value. When
discussing the early time evolution of $\Phi$ various effects
need to be considered \cite{displacement,fvi,bertbento}, but the outcome
\cite{ours} is that at the epoch $H\sim m_\Phi$ it starts to
oscillate about $\Phi\simeq 0$ with amplitude of order $m_{\rm Pl}$.
The corresponding abundance
$ n_\Phi / s \sim ( m_{\rm Pl} / m_\Phi )^{1/2} \sim 10^8$
is cosmologically insignificant, bearing in mind the dilution factor.
However, at the end of Thermal Inflation
$\Phi$ is in general still displaced from its vacuum value by
its interaction with the GUT Higgs, and by an amount which turns out to
be large compared with the oscillation.
To estimate this displacement \cite{ours}, recall that the quantity $V_0$
appearing in Eq.~(\ref{V}) is supposed to be generated dynamically from
the Planck scale $m_{\rm Pl}$. In that equation $\Phi=0$,
but for fixed $|\Phi| \ll m_{\rm Pl}$ a similar equation will hold
with some $V_0(\Phi)$, whose slope $ \partial V_0 / \partial \Phi $
will be of order $V_0/m_{\rm Pl}$.
The effective potential for $\Phi$ in the regime $|\Phi| \ll m_{\rm Pl}$ is
then $ m_{\Phi}^2 \Phi^2 /2 + (\partial V_0/\partial \Phi)\Phi $,
so the displacement is of order
$ V_0 m_{\Phi}^{-2} m_{\rm Pl}^{-1} \sim (H/m_\Phi)^2 m_{\rm Pl}$.
If after Thermal Inflation the effective potential promptly reverted to
$ m_\Phi^2 \Phi^2 / 2 $, then $\Phi$ would start to oscillate with this
amplitude corresponding to
$ n_\Phi \sim V_{0}^{2} m_{\Phi}^{-3} m_{\rm Pl}^{-2} $.
In that case the abundance at nucleosynthesis would be
\begin{eqnarray}
\frac{n_\Phi}{s} & \sim & 10^{-5} \left(\frac{M_{\rm GUT} }{m_{\rm Pl}}
\right)^2
	\times \nonumber\\
 & & \left(\frac{ T_{\rm decay} }{ 10\,{\rm MeV} }\right)
\left( \frac{ V_0 }{ m_\Phi^2 M_{\rm GUT}^2 }\right)
\left( \frac{ 1\,{\rm TeV} }{ m_\Phi } \right)
\end{eqnarray}
The first line is of order $10^{-10}$ and the remaining factors are
roughly of order 1. In reality the dynamics at the end of Thermal
Inflation will be quite complicated but this estimate should
still be reasonable \cite{ours}. Taking into account the considerable
uncertainty, the conclusion is that the moduli problem may be solved.

The classical displacement discussed in the last paragraph
was not taken into account by Randall and Thomas \cite{Randall}
when they claimed that the moduli problem
can be solved by several $e$-folds of inflation at the scale
$ V_{\rm inf} \sim m_{\Phi}^2 m_{\rm Pl}^2 $.
{}From the above discussion one in fact needs
$ V_{\rm inf} \lesssim
10^{-7} ( 10^{12} A ) ( 10\,{\rm MeV} / T_{\rm decay} )
( m_\Phi / 1\,{\rm TeV} ) m_{\Phi}^2 m_{\rm Pl}^2 $
to solve the moduli problem in this way, where $A$ is the bound on
$ n_\Phi / s $ at nucleosynthesis.

Stable topological defects may be produced at the end of the first era
of inflation, and at the GUT transition.
Let us look briefly at the case of gauge monopoles
produced at the GUT transition,
and assume that they are not connected by strings.
The strongest bound on their abundance comes from baryon
decay catalysis in neutron stars, which gives
$n/s\lesssim 10^{-37}$ \cite{frieman}.
The temperature is too low for annihilation \cite{vs}, but one
monopole per Hubble volume at creation gives
$n/s\sim 10^{-3}(M_{\rm GUT}/m_{\rm Pl})^3\sim 10^{-10}$,
which requires a dilution
factor $\Delta\sim 10^{27}$. Thus there may be no monopole problem.

So much for undesirable relics. What about desirable ones, in the form
of matter? Hot Dark Matter (massive neutrinos) has the usual abundance
because its freeze-out temperature is a few MeV and hence less than
$T_{\rm decay}$. If it is stable, the LSP (lightest supersymmetric particle)
will be Cold Dark Matter. It is not produced after GUT Higgs decay
in our cosmology
because its freeze-out temperature is of order
$1\,\mbox{GeV} $, but it will be produced
by the GUT Higgs decay unless $m_h < 2 m_{\rm LSP}$.
If $N$ LSP's are produced per GUT Higgs, then
$ n_{\rm LSP}/s \sim N (T_{\rm decay}/m) $ and
$ \Omega_{\rm LSP} \sim 10^{10} N (T_{\rm decay}/m)
(m_{\rm LSP}/10\,\mbox{GeV})
\sim 10^5 N (m_{\rm LSP}/10\,\mbox{GeV}) $.
Since $N<10^{-5}$ seems unlikely we probably need either
$R$-parity violation to destabilize the LSP, or
$m_h < 2 m_{\rm LSP}$.

For baryogenesis, the most commonly considered mechanisms in the
usual cosmology are \cite{Dolgov}
non-perturbative effects at the electroweak transition,
particle decay and the Affleck-Dine mechanism \cite{affleckdine}.
In our cosmology the electroweak and GUT transitions happen at more or
less the same time, but without going into detail it seems clear that the
first mechanism cannot be significant because of the dilution factor.
However, if $R$-parity is violated the baryons might be created
in the GUT Higgs decay \cite{yam2,dimshafi}.
The Affleck-Dine mechanism can generate both baryons and the LSP after
Thermal Inflation \cite{ours}.

The other favoured Cold Dark Matter candidate is the axion.
Axion cosmology is quite subtle \cite{inflation,vs,ouraxion,paul}.
For simplicity let us ignore the saxino
and axino (the axion's superpartners).
Recall that the axion field is $a=f_{\rm a}\theta$ where $\theta$ is
the `misalignment angle' and $f_{\rm a}$ is related to the mass by
$ m_{\rm a} / 10^{-3}\,\mbox{eV}
= 0.62 \times 10^{10}\,\mbox{GeV} / f_{\rm a} $.
{}From accelerator physics and astrophysics,
$m_{\rm a}\lesssim 10^{-2} \,\mbox{eV}$.
The axion mass switches on gradually
as $T$ falls towards $100\,{\rm MeV}$.

Let us first suppose that there are no axionic strings.
Then $\theta$ is typically homogeneous with some initial value
$\tilde \theta$, and in the standard cosmology it
starts to oscillate when $m_{\rm a}(T)\sim H$, leading to
$\Omega_{\rm a}\sim \tilde\theta^2 (10^{-5}\,\mbox{eV}/m_{\rm a})^{1.2}$.
In our cosmology oscillation starts when $m_{\rm a}\sim H$, the temperature
being negligible, and this leads to $\Omega_{\rm a}\sim \tilde\theta^2
(10^{-8}\,\mbox{eV}/m_{\rm a})^2 (T_{\rm decay}/10\,{\rm MeV})$.

Now suppose that there are strings.
In the standard cosmology there is roughly one string per Hubble volume,
until $m_{\rm a}(T) \sim H$ when domain walls form between the strings
and the wall/string network annihilates, and axions radiated from strings
prior to this epoch give \cite{paul}
$\Omega_{\rm a} \sim (10^{-3}\,\mbox{eV}/m_{\rm a})$.
In our cosmology the string spacing leaves the horizon at the beginning of
Thermal Inflation, and the axion field then freezes until
$H \sim m_{\rm a}$.
At that epoch domain walls form, and between them the almost
homogeneous axion field oscillates to give a contribution
$ \Omega_{\rm a} \sim (10^{-8}\,\mbox{eV}/m_{\rm a})^2
(T_{\rm decay}/10\,{\rm MeV})$.
The string-wall network re-enters the horizon at the epoch
$\rho_{\rm entry}^{1/4} \sim 10m(m/M_{\rm GUT})^{1/4}$ corresponding to
$ \rho_{\rm entry}^{1/4} / T_{\rm decay} \sim
10^3 (M_{\rm GUT}/10^{16}\,{\rm GeV})^{3/4} $, when it decays into
marginally relativistic axions giving a contribution
$\Omega_{\rm a} \sim
(10^{-8}\,\mbox{eV}/m_{\rm a}) (10^{16}\,{\rm GeV}/M_{\rm GUT})^{1/2}$.
The overall conclusion is that the axions can provide Cold Dark Matter
in our cosmology, provided that $m_{\rm a} \lesssim 10^{-8}\,{\rm eV}$
corresponding to $f_{\rm a} \gtrsim 10^{15}\,{\rm GeV}$.

Except in the last paragraph we have ignored the inhomogeneity of the
universe. One might wonder if GUT Higgs `stars' could form during the
Cold Big Bang (cf. \cite{Steinhardt}). If they form sufficiently early
they might be dense enough to briefly thermalize the GUT Higgs decay
products. During ordinary inflation the quantum fluctuation effectively
generates a classical curvature perturbation as each scale leaves the
horizon, which remains constant until horizon re-entry and has a roughly
scale-independent magnitude $\lesssim 10^{-5}$. One does not expect a
significant quantum fluctuation during Thermal Inflation because
$H \ll m$. If the gap between ordinary and Thermal inflation is
negligible, this information allows one to
estimate the density perturbation.
On the scale leaving the horizon at the
beginning of Thermal Inflation it is
$\lesssim 10^{-5}$ at the epoch of re-entry, and it then grows like
$\rho^{-1/3}$ to become $\lesssim 10^{-1}$ at GUT Higgs decay.
On bigger scales it is smaller, and on smaller
scales it vanishes, so there is no significant structure formation.
To extend this analysis to the case where there is a gap
one would have to consider the evolution of the curvature
perturbation inside the horizon during Thermal Inflation.

Assuming a flat GUT Higgs potential, the main alternative to our
cosmology would be to have $|h|$ initially displaced from 0,
so that at the epoch $H\sim m$ it starts to oscillate
about its vacuum value \cite{Ross}. The GUT Higgs particles
produced in this way must still decay before nucleosynthesis, and we do
not now solve the moduli problem. The other alternative, which we have
not considered, would be to retain the initial value $|h|=0$, but to
relax the assumption that $T_{\rm R} \gtrsim V_0^{1/4}$.

In this article we have taken the scale of symmetry breaking to
be $10^{16}\,\mbox{GeV}$, because this is what experiment indicates for a
gauge symmetry.
{}From the point of view of cosmology an attractive scale is
$\sim 10^{13}\,\mbox{GeV}$,
because it minimizes the moduli abundance making
the nucleosynthesis constraint easier to satisfy,
and it also gives $T_{\rm decay} \sim 1\,\mbox{GeV}$ which might be
high enough for the LSP to thermalise.
Such a scale might be associated with the breaking of a global
symmetry, and our cosmology could work equally well in that case.

\underline{Acknowledgements}: EDS is supported by the Royal Society,
and we both acknowledge support from the Newton Institute, Cambridge,
where this work was begun. We thank Ed Copeland, Andrew Liddle,
Tomislav Prokopec and Subir Sarkar for useful discussions.

\begin{table}
\centering
\caption[History of the Early Universe]
{History of the Early Universe.
There are large uncertainties in our estimates.}
\begin{tabular}{ c c c l}
$\rho^{1/4}$ & $T$ & $H$ & \\
\hline
$V_{\rm inf}^{1/4}$ & 0 & $V_{\rm inf}^{1/2}/m_{\rm Pl}$
& Ordinary inflation ends\\
$T_{\rm R}$ & $T_{\rm R}$ & $T_{\rm R}^2/m_{\rm Pl}$ & Hot Big Bang begins \\
&&& (unbroken GUT vacuum) \\
$10^9\,\mbox{GeV}$ & $10^9\,\mbox{GeV}$ & $100\,{\rm MeV}$
& Thermal Inflation begins\\
$10^9\,\mbox{GeV}$ & $10^3\,\mbox{GeV}$ & $100\,{\rm MeV}$
& Cold Big Bang begins \\
&&& (present day vacuum) \\
$10\,{\rm MeV}$ & $10^{-3}\,\mbox{eV}$ & $10^{-14}\,\mbox{eV}$
& GUT Higgs decay starts \\
$10\,{\rm MeV}$ & $10\,{\rm MeV}$ & $10^{-14}\,\mbox{eV}$
& Hot Big Bang begins \\
&&& (present day vacuum)\\
\end{tabular}
\end{table}

\end{document}